\documentclass[preprint,showpacs,preprintnumbers,amsmath,amssymb]{revtex4}

\usepackage{graphicx}
\usepackage{dcolumn}
\usepackage{bm}

\usepackage{amssymb}
\usepackage{latexsym}
\usepackage{amsthm}
\usepackage{graphicx}
\theoremstyle{plain}

\newtheorem{theorem}{Theorem}

\newtheorem{corollary}{Corollary}
\newtheorem{proposition}{Proposition}

\newcommand{\RM}{\mathbb{R}}
\newcommand{\ZM}{\mathbb{Z}}

\newcommand{\TM}{\mathbb{T}}

\newcommand{\tpm}{\TM^{(p)} _M}
\newcommand{\mpi}{\mu^{(p)} _{\infty} (x)}
\newcommand{\mpm}{\mu^{(p)} _{M} (x)}

\begin{document}

\title{Continuous-Time Quantum Walks on Trees \\ in Quantum Probability Theory}

\author{Norio Konno}
\email{konno@ynu.ac.jp}
\affiliation{%
Department of Applied Mathematics, 
Yokohama National University, 
79-5 Tokiwadai, Yokohama, 240-8501, Japan\\}




\begin{abstract}
A quantum central limit theorem for a continuous-time quantum walk on a homogeneous tree is derived from quantum probability theory. As a consequence, a new type of limit theorems for another continuous-time walk introduced by the walk is presented. The limit density is similar to that given by a continuous-time quantum walk on the one-dimensional lattice.
\end{abstract}

\maketitle



\section{Introduction}

Two types of quantum walks, discrete-time or continuous-time, were introduced as the quantum mechanical extension of the corresponding classical random walks and have been extensively studied over the last few years, see \cite{kem,tre} for recent reviews. In this paper we consider a continuous-time quantum walk on a homogeneous tree in quantum probability theory. The walk is defined by identifying the Hamiltonian of the system with a matrix related to the adjacency matrix of the tree. Concerning continuous-time quantum walks, see \cite{chi02,chi03,ger,ahm,ada,ben,inu,kon05,got,sol,fed,jaf,mul,mul06} for examples.

Let $\tpm$ denote a homogeneous tree of degree $p$ with $M$-generation. After we fix a root $o \in \tpm$, a stratification (distance partition) is introduced by the natural distance function in the following way:
\begin{eqnarray*}
\tpm = \bigcup_{k=0} ^M V_{k} ^{(p)}, \quad V_{k} ^{(p)} = \{ x \in \tpm : \partial (o,x)=k \}.
\end{eqnarray*}
Here $\partial (x,y)$ stands for the length of the shortest path connecting $x$ and $y$. Then
\begin{eqnarray*}
|V_0 ^{(p)}|=1, \> |V_1 ^{(p)}|=p,  \> |V_2 ^{(p)}|=p(p-1), \ldots, |V_k ^{(p)}|=p(p-1)^{k-1}, \ldots,
\end{eqnarray*}
where $|A|$ is the number of elements in a set $A$. 
The total number of points in $M$-generation, $ |\tpm |,$ is $p(p-1)^M - (p-1).$

Let $H_M ^{(p)}$ be a $|\tpm| \times |\tpm|$ symmetric matrix given by the adjacency matrix of the tree $\tpm.$ The matrix is treated as the Hamiltonian of the quantum system. The $(i,j)$ component of $H_M ^{(p)}$ denotes $H_M ^{(p)}(i,j)$  for $i,j \in \{0,1, \ldots, |\tpm|-1 \}.$ In our case, the diagonal component of $H_M ^{(p)}$ is always zero, i.e., $H_M ^{(p)}(i,i) = 0$ for any $i$. On the other hand, the diagonal component of corresponding matrix $H_{M,MB} ^{(p)}$ investigated in \cite{mul} for $p=3$ case is not zero. For example, $H_{1,MB} ^{(3)}(0,0) = -3, \> H_{1,MB} ^{(3)}(1,1) = H_{1,MB} ^{(3)}(2,2) = H_{1,MB} ^{(3)}(3,3) = -1.$

The evolution of continuous-time quantum walk on the tree of $M$-generation, $\tpm$, is governed by the following unitary matrix:
\begin{eqnarray*}
U^{(p)} _M (t) = e^{it H_M ^{(p)}}.
\label{eqn:unitary}
\end{eqnarray*}

The amplitude wave function at time $t$, $| \Psi_M ^{(p)} (t) \rangle $, is defined by 
\begin{eqnarray*}
| \Psi_M ^{(p)} (t) \rangle = U_M ^{(p)} (t) | \Psi_M ^{(p)} (0) \rangle . 
\label{eqn:evolution}
\end{eqnarray*}
In this paper we take $| \Psi_M ^{(p)} (0) \rangle = [1,0,0, \ldots, 0]^T$ as an initial state, where $T$ denotes the transposed operator.

The $(n+1)$-th coordinate of $| \Psi_M ^{(p)} (t) \rangle$ is denoted by $| \Psi_M ^{(p)} (n,t) \rangle$ which is the amplitude wave function at site $n$ at time $t$ for $n=0,1, \ldots , p(p-1)^M -p.$ The probability finding the walker is at site $n$ at time $t$ on $\tpm$ is given by
\begin{eqnarray*}
P_M ^{(p)} (n,t) = \langle \Psi_M ^{(p)} (n,t) | \Psi_M ^{(p)} (n,t) \rangle .
\label{eqn:prob}
\end{eqnarray*}
Then we define the continuous-time quantum walk $X^{(p)} _M (t)$ at time $t$ on $\tpm$ by 
\begin{eqnarray*}
P( X^{(p)} _M (t) = n) = P_M ^{(p)} (n,t) .
\label{eqn:prob}
\end{eqnarray*}
In a similar way, let $X^{(p)} _{M,MB} (t)$ be a quantum walk given by $H_{M,MB} ^{(p)}$. As we stated before, $H_{M,MB} ^{(p)} (i,i)$ depends on $i$ for any finite $M$. However in $M \to \infty$ limit, the $(i,i)$ component of the matrix becomes $-p$ for any $i$. Remark that the probability distribution of the continuous-time walk does not depend on the value of the diagonal component of the scalar matrix. Therefore the definitions of the walks imply that both quantum walks coincide in $M \to \infty$ limit, i.e.,  
\begin{eqnarray*}
\lim_{M \to \infty} P( X^{(p)} _M (t) = n) = \lim_{M \to \infty} P( X^{(p)} _{M,MB} (t) = n),
\label{eqn:prob}
\end{eqnarray*}
for any $t$ and $n$.

This paper is organized as follows. In Sec. 2, we review the quantum probabilistic approach and give preliminaries and some examples for the walk $X^{(p)} _M (t)$. A quantum central limit theorem as $p \to \infty$ is derived from quantum probability theory in Sec. 3. Finally we present a limit theorem for another continuous-time walk $Y(t)$ introduced as a $p$-limit walk of $X^{(p)} _{\infty} (t)$.

\section{Quantum Probabilistic Approach}

\subsection{Finite $M$ case}

Let $\mu^{(p)} _M$ denote the spectral distribution of our adjacency matrix $H_{M} ^{(p)}$. From the general theory of an interacting Fock space (see \cite{aca,has,oba04,iga,oba,jaf}, for examples), the orthogonal polynomials $\{ Q^{(p)} _n \}$ and $\{ Q^{(p,\ast)} _n \}$ associated with $\mu^{(p)} _M$ satisfy the following three-term recurrence relations with a Szeg\"o-Jacobi parameter ($\{ \omega_n \}, \{ \alpha_n \}$) respectively:
\begin{eqnarray*}
&&
Q_0 ^{(p)} (x) =1, \>\> Q_1 ^{(p)}(x) =x - \alpha_1 , \>\> 
\\
&&
x Q_n ^{(p)} (x) = Q_{n+1} ^{(p)} (x) + \alpha_{n+1} Q_n ^{(p)} (x) + \omega_{n} Q_{n-1} ^{(p)} (x) \>\> (n \ge 1),
\end{eqnarray*}
and
\begin{eqnarray*}
&&
Q_0 ^{(p,\ast)} (x) =1, \>\> Q_1 ^{(p),\ast}(x) =x - \alpha_2, \>\> 
\\
&&
x Q_n ^{(p,\ast)} (x) = Q_{n+1} ^{(p,\ast)} (x) + \alpha_{n+2} Q_n ^{(p,\ast)} (x) + \omega_{n+1} Q_{n-1} ^{(p,\ast)} (x) \>\> (n \ge 1).
\end{eqnarray*}
In our tree case, 
\begin{eqnarray*}
\omega_1=p, \> \omega_2 = \omega_3 = \cdots = \omega_M = p-1, \> \omega_{M+1}= \omega_{M+2}= \cdots =0, \>\> \alpha_1= \alpha_2= \cdots =0.
\end{eqnarray*}
Then the Stieltjes transform $G_{\mu^{(p)} _M}$ of $\mu^{(p)} _M$ is given by
\begin{eqnarray*}
G_{\mu^{(p)} _M} (x) = {Q_{n-1} ^{(p,\ast)} (x) \over Q_{n} ^{(p)} (x)},
\end{eqnarray*}
where $n=|\tpm | = p(p-1)^M - (p-1).$ 

The following result was shown in \cite{jaf}:
\begin{eqnarray*}
| \Psi_{M} ^{(p)} (V_k ^{(p)},t) \rangle = 
{ 1 \over \sqrt{|V_k ^{(p)}|} } \> \int_{\RM} \> \exp (itx) \> Q_k ^{(p)} \mpm \> dx,
\end{eqnarray*}
for $k = 0,1,2, \ldots.$ Remark that $|V_k ^{(p)}| = \omega_1 \omega_2 \cdots \omega_k = p(p-1)^{k-1} \> (1 \le k \le M)$ and $|V_0 ^{(p)}|=1.$ It is important to note that
\begin{eqnarray*}
| \Psi_{M} ^{(p)} (n,t) \rangle
= { 1 \over |V_k ^{(p)}| } \> \int_{\RM} \> \exp (itx) \> Q_k ^{(p)} (x) \mpm \> dx \quad \hbox{if} \>\> n \in V_k ^{(p)} \>\> (k=0,1, \ldots, M).
\end{eqnarray*}
The proof appeared in Appendix A in \cite{jaf}.

\subsection{$p=3$ and $M=2$ case}

Here we consider $p=3$ and $M=2$ case. Then we have $n=10,\> \omega_1=3, \omega_2 =2, \omega_3= \omega_4= \cdots =0, \> \alpha_1= \alpha_2= \cdots =0.$ The definitions of $Q^{(3)} _n (x)$ and $Q_n ^{(3,\ast)} (x)$ imply  
\begin{eqnarray*}
Q_0 ^{(3)} (x) =1, \>\> Q_1 ^{(3)}(x) =x, \>\> Q_2 ^{(3)} (x) = x^2 - 3, \>\> Q_k ^{(3)} (x) =x^{k-2} (x^2-5) \>\> (k \ge 3),
\end{eqnarray*}
and
\begin{eqnarray*}
Q_0 ^{(3,\ast)} (x) =1, \>\> Q_1 ^{(3,\ast)} (x) =x, \>\> Q_k ^{(3,\ast)} (x) =x^{k-2} (x^2-2)  \>\> (k \ge 2).
\end{eqnarray*}
Therefore we obtain the Stieltjes transform:
\begin{eqnarray*}
G_{\mu^{(3)} _2} (x) = {Q_9 ^{(3,\ast)} (x) \over Q_{10} ^{(3)} (x)}
= {2 \over 5} \cdot {1 \over x} + {3 \over 10} \cdot {1 \over x - \sqrt{5}} + {3 \over 10} \cdot {1 \over x + \sqrt{5}}.
\end{eqnarray*}
From this, we see that
\begin{eqnarray*}
\mu^{(3)} _2  = {2 \over 5} \> \delta_0 (x) + {3 \over 10} \> \delta_{- \sqrt{5}} (x) + {3 \over 10} \> \delta_{\sqrt{5}} (x). 
\end{eqnarray*}
Then
\begin{eqnarray*}
| \Psi_{2} ^{(3)} (V_0 ^{(3)},t) \rangle 
&=& \int_{\RM} \> \exp (itx) \> \mu^{(3)} _2 (dx)
= {1 \over 5} \> (2 + 3 \cos (\sqrt{5} t) ),
\\
| \Psi_{2} ^{(3)} (V_1 ^{(3)},t) \rangle 
&=& {1 \over \sqrt{\omega_1}} \> \int_{\RM} \> \exp (itx) Q_1 ^{(3)} (x) \> \mu^{(3)} _2 (dx)
= {i \sqrt{3} \over \sqrt{5}} \> \sin (\sqrt{5} t),
\\
| \Psi_{2} ^{(3)} (V_2 ^{(3)},t) \rangle 
&=& {1 \over \sqrt{\omega_1 \omega_2}} \> \int_{\RM} \> \exp (itx) Q_2 ^{(3)} (x) \> \mu^{(3)} _2 (dx)
= {\sqrt{6} \over 5} \left( - 1 +  \cos (\sqrt{5} t) \right).
\end{eqnarray*}
Noting that $| \Psi_{2} ^{(3)} (n,t) \rangle = | \Psi_{2} ^{(3)} (V_k ^{(3)},t) \rangle / \sqrt{|V_k ^{(3)}|}$ for any $k=0,1,2$, we obtain the same conclusion as the result given by the eigenvalues and the eigenvectors of $H^{(3)} _2.$

\subsection{$M \to \infty$ case}

The quantum probabilistic approach \cite{iga,oba,jaf} implies that
\begin{eqnarray*}
| \Psi_{\infty} ^{(p)} (V_k ^{(p)},t) \rangle = \lim_{M \to \infty} |\Psi_{M} ^{(p)} (V_k,t) \rangle = { 1 \over \sqrt{|V_k ^{(p)}|} } \> \int_{\RM} \> \exp (itx) \> Q_k ^{(p)} (x) \mpi \> dx, 
\end{eqnarray*}
for $k=0,1,2, \ldots$, where the limit spectral distribution $\mpi$ is given by
\begin{eqnarray*}
I_{(-2 \sqrt{p-1}, 2 \sqrt{p-1})} (x) \> {p \sqrt{4(p-1)-x^2} \over 2 \pi (p^2 - x^2)}.
\end{eqnarray*}
Here $I_A$ is the indicator function of $A$, i.e., $I_A (x) = 1,$ if $x \in A, \> =0,$ if $x \not\in A.$ This type of measure was first obtained by Kesten \cite{kes} in a classical random walk with a different method. An immediate consequence is 
\begin{eqnarray*}
P_{\infty} ^{(p)} (V_k ^{(p)},t) = { 1 \over |V_k ^{(p)}| } \> \left[ 
\left\{ \int_{\RM} \> \cos (tx) \> Q_k ^{(p)} (x) \mpi \> dx \right\}^2 +
\left\{ \int_{\RM} \> \sin (tx) \> Q_k ^{(p)} (x) \mpi \> dx \right\}^2 
\right], 
\end{eqnarray*}
for $k=0,1,2, \ldots $ Furthermore, as in the case of finite $M$, we see that
\begin{eqnarray}
| \Psi_{\infty} ^{(p)} (n,t) \rangle
= { 1 \over |V_k ^{(p)}| } \> \int_{\RM} \> \exp (itx) \> Q_k ^{(p)} (x) \mpi \> dx,
\label{eqn:pocky}
\end{eqnarray}
if $n \in V_k ^{(p)} \>\> (k=0,1,2, \ldots).$ From (\ref{eqn:pocky}) and the Riemann-Lebesgue lemma, we have $\lim_{t \to \infty} |\Psi_{\infty} ^{(p)} (n,t) \rangle = 0,$ for any $n$, since $ Q_k ^{(p)} (x) \mu^{(p)} _{\infty} (x) \in L^1 (\RM).$ Therefore we see that $\lim_{t \to \infty} P_{\infty} ^{(p)} (n,t)  = 0.$ So we conclude that $\bar{P}_{\infty} ^{(p)} (n)  = 0,$ where $\bar{P}_{\infty} ^{(p)} (n)$ is the time-averaged distribution of $P_{\infty} ^{(p)} (n,t).$

\subsection{$p=2$ and $M \to \infty$ case}

In this subsection, we consider $p=2$ and $M \to \infty$, i.e., $\ZM^1$ case. Then we have 

\begin{proposition}
\begin{eqnarray*}
| \Psi_{\infty} ^{(2)} (V_0 ^{(2)},t) \rangle
= J_0 (2t),
\qquad 
| \Psi_{\infty} ^{(2)} (V_k ^{(2)},t) \rangle
= 
\sqrt{2} \> i^k \> J_k (2t),
\quad
(k=0,1,2, \ldots),
\end{eqnarray*}
where $J_n (x)$ is the Bessel function of the first kind of order $n$. 
\end{proposition}
\begin{proof}
We induct on $k$. For $k=0$ case, we use the following result (see (4) in page 48 in \cite{wat}):
\begin{eqnarray}
\int_{- 1} ^{1} \> \exp \left( i s x \right) \> 
(1 - x^2)^{\nu - 1/2} \> dx 
= {\Gamma (1/2) \Gamma (\nu + 1/2) \over (s/2)^{\nu} } \> J_{\nu} (s),
\label{eqn:trino}
\end{eqnarray}
where $\Gamma (x)$ is the Gamma function. Combining $\Gamma (3/2) = \sqrt{\pi}/2, \> \Gamma (1/2) = \sqrt{\pi}$ with $Q^{(2)}_0 (x)=1$ and $\nu = 0$ gives
\begin{eqnarray*}
| \Psi_{\infty} ^{(2)} (V_0 ^{(2)},t) \rangle
= \int_{- 2} ^{2} \> \exp \left( i t x \right) \> {1 \over \pi \sqrt{4 - x^2}} \> dx 
\> 
=
\>
J_0 (2t).
\end{eqnarray*}
In a similar fashion, we verify that the result holds for $k=1,2.$

Next we suppose that the result is true for all values up to $k$, where $k \ge 2$. Then we see that 
\begin{eqnarray*}
&&
| \Psi_{\infty} ^{(2)} (V_{k+1} ^{(2)},t) \rangle
\\
&=& {1 \over \sqrt{2}} \> \int_{- 2} ^{2} \> \exp \left( i t x \right) \> Q^{(2)} _{k+1} (x) \> {dx \over \pi \sqrt{4 - x^2}} 
\\
&=&
{1 \over \sqrt{2}} \> \int_{- 2} ^{2} \> \exp \left( i t x \right) \> \left\{ x Q^{(2)} _{k} (x) -  Q^{(2)} _{k-1} (x) \right\} \> {dx \over \pi \sqrt{4 - x^2}}
\\
&=&
{1 \over i} \> {d \over dt} \left( {1 \over \sqrt{2}} \> \int_{- 2} ^{2} \> \exp \left( i t x \right) \>  Q^{(2)} _{k} (x) \> {dx \over \pi \sqrt{4 - x^2}} \right) -  {1 \over \sqrt{2}} \> \int_{- 2} ^{2} \> \exp \left( i t x \right) \> Q^{(2)} _{k-1} (x) \> {dx \over \pi \sqrt{4 - x^2}}
\\
&=&
{1 \over i} \> {d \over dt} (\sqrt{2} \> i^k \> J_k (2t)) - \sqrt{2} \> i^{k-1} \> J_{k-1} (2t)
\\
&=&
\sqrt{2} \> i^{k+1} \> J_{k+1} (2t).
\end{eqnarray*}
The second equality follows from the definition of $Q^{(2)} _{k} (x).$ By induction, we have the fourth equality. For the last equality, we use a recurrence formula for the Bessel coefficients: $2 J_{k}'(2t)=J_{k-1} (2t) - J_{k+1}(2t)$ (see (2) in page 17 of \cite{wat}). 
\end{proof}
As a consequence, we have
\begin{corollary}
\begin{eqnarray*}
P_{\infty} ^{(2)} (V_0 ^{(2)},t)
= J_0^2 (2t),
\qquad
P_{\infty} ^{(2)} (V_k ^{(2)},t) 
= 
2 J_k ^2 (2t),
\quad
(k=1,2, \ldots).
\end{eqnarray*}
\end{corollary}
We confirm that 
\begin{eqnarray*}
\sum_{k=0} ^{\infty} P_{\infty} ^{(2)} (V_k ^{(2)},t) 
=1, 
\end{eqnarray*}
since it follows from $J_0 ^2 (2t) + 2 \sum_{k=1} ^{\infty} J_k ^2 (2t) =1$ (see (3) in page 31 in \cite{wat}). Noting that $V_k ^{(2)} = \{ -k, k \}$ for any $k \ge 0$, we have the same result given by \cite{kon05}:
\begin{eqnarray*}
P_{\infty} ^{(2)} (n,t) =  J_n ^2 (2t),
\end{eqnarray*}
for any $n \in \ZM$ and $t \ge 0$.

\section{Quantum Central Limit Theorem}

To state a quantum central limit theorem in our case, it is convenient to rewrite as 
\begin{eqnarray*}
\left\langle \Phi^{(p)} _k \left| \exp \left( i t H^{(p)} _{\infty}  \right) \right| \Phi^{(p)} _0 \right\rangle 
= | \Psi_{\infty} ^{(p)} (V_k ^{(p)},t) \rangle,
\end{eqnarray*}
where
\begin{eqnarray*}
\Phi^{(p)} _k = { 1 \over \sqrt{|V_k ^{(p)}|} } \> \sum_{n \in V_k ^{(p)}} I_n,
\end{eqnarray*}
and $I_n$ denotes the indicator function of the singleton $\{ n \}.$ It is easily obtained that
\begin{eqnarray*}
\lim_{p \to \infty} \left\langle \Phi^{(p)} _k \left| \exp \left( i t H^{(p)} _{\infty}  \right) \right| \Phi^{(p)} _0 \right\rangle 
= 0,
\end{eqnarray*}
for any $k \ge 0.$ Then we have the following quantum central limit theorem:

\begin{theorem} 

\begin{eqnarray*}
\lim_{p \to \infty} \left\langle \Phi^{(p)} _k \left| \exp \left( i t {H^{(p)} _{\infty} \over \sqrt{p}}  \right) \right| \Phi^{(p)} _0 \right\rangle 
=  (k+1) \> i^k \> {J_{k+1} (2t) \over t}, 
\end{eqnarray*}
for $k=0,1,2, \ldots.$

\end{theorem}

\begin{proof} We induct on $k$. First we consider $k=0$ case. We see that
\begin{eqnarray*}
&& 
\lim_{p \to \infty} 
\left\langle \Phi^{(p)} _0 \left| \exp \left( i t { H^{(p)} _{\infty} \over \sqrt{p} } \right) \right| \Phi^{(p)} _0 \right\rangle 
\\
&=&
\lim_{p \to \infty} 
\int_{\RM} \> \exp \left(it {x \over \sqrt{p} } \right) \> \mpi \> dx 
\\
&=& 
\lim_{p \to \infty} 
\int_{- 2 \sqrt{(p-1)/p}} ^{2 \sqrt{(p-1)/p}} \> \exp \left( it x \right) \> 
{\sqrt{(2(p-1)/p)^2 - x^2} \over 2 \pi ( 1 - x^2/p)} \> dx
\\
&=&
\int_{- 1} ^{1} \> \exp \left( 2 i t x \right) \> 
{2 \sqrt{1 - x^2} \over \pi} \> dx
\end{eqnarray*}
Then (\ref{eqn:trino}) with $\nu = 1$ yields
\begin{eqnarray*}
\lim_{p \to \infty} \left\langle \Phi^{(p)} _0 \left| \exp \left( i t { H^{(p)} _{\infty} \over \sqrt{p} } \right) \right| \Phi^{(p)} _0 \right\rangle = {J_1 (2t) \over t}.
\end{eqnarray*}
So the result holds for $k=0$. Similarly we obtain  
\begin{eqnarray*}
\lim_{p \to \infty} \left\langle \Phi^{(p)} _1 \left| \exp \left( i t {H^{(p)} _{\infty} \over \sqrt{p}}  \right) \right| \Phi^{(p)} _0 \right\rangle 
&=& {2 i J_2 (2t) \over t},
\\ 
\lim_{p \to \infty} \left\langle \Phi^{(p)} _2 \left| \exp \left( i t {H^{(p)} _{\infty} \over \sqrt{p}}  \right) \right| \Phi^{(p)} _0 \right\rangle 
&=& -{3 J_3 (2t) \over t}.
\end{eqnarray*}
Next we suppose that the result holds for all values up to $k$, where $k \ge 2$. Then we have 
\begin{eqnarray*}
&&
\lim_{p \to \infty} 
\left\langle \Phi^{(p)} _{k+1} \left| \exp \left( i t {H^{(p)} _{\infty} \over \sqrt{p}}  \right) \right| \Phi^{(p)} _0 \right\rangle 
\\
&=& 
\lim_{p \to \infty} 
{1 \over \sqrt{|V_{k+1} ^{(p)}|}} \int_{\RM} \> \exp \left( it {x \over \sqrt{p}} \right) \> Q_{k+1} ^{(p)} (x) \> \mpi \> dx 
\\
&=& 
\lim_{p \to \infty} 
{1 \over \sqrt{ p(p-1)^{k}}} 
\int_{- 2 \sqrt{(p-1)/p}} ^{2 \sqrt{(p-1)/p}} \> \exp \left( it x \right) 
\> Q_{k+1} ^{(p)} (\sqrt{p} x) \> {\sqrt{(2(p-1)/p)^2 - x^2} \over 2 \pi ( 1 - x^2/p)} \> dx
\\
&=& 
\int_{- 2} ^{2} \> \exp \left( it x \right) \> 
Q^{(\infty)}_{k+1} (x) \>
{\sqrt{2^2 - x^2} \over 2 \pi} \> dx
\\
&=&
\int_{- 2} ^{2} \> \exp \left( it x \right) \> 
\left\{ x Q^{(\infty)}_{k} (x) - Q^{(\infty)}_{k-1} (x) \right\} \>
{\sqrt{2^2 - x^2} \over 2 \pi} \> dx
\\
&=&
{1 \over i} {d \over dt} \left( 
\int_{- 1} ^{1} \> \exp \left( 2 i t x \right) \> 
Q^{(\infty)}_k (2x) \> {2 \sqrt{1 - x^2} \over \pi} \> dx \right)
\\
&&
\qquad \qquad \qquad - 
\int_{- 1} ^{1} \> \exp \left( 2 i t x \right) \> 
Q^{(\infty)}_{k-1} (2x) \> {2 \sqrt{1 - x^2} \over \pi} \> dx
\\
&=&
i^{k-1} \left\{ (k+1) {d \over dt} \left( J_{k+1} (2t) \over t \right) 
- k {J_k (2t) \over t } \right\}
\end{eqnarray*}
where the last equality is given by the induction and $Q^{(\infty)}_k (x) = \lim_{p \to \infty} Q^{(p)} _k (\sqrt{p} x) / \sqrt{ p(p-1)^{k-1}},$ if the right hand side exists. We confirm that the limit exists for any $k \ge 1$. For example, we compute $Q^{(\infty)}_1 (x) =x, \> Q^{(\infty)}_2 (x) =x^2-1, \> Q^{(\infty)}_3 (x) = x^3 -2 x, \> Q^{(\infty)}_4 (x) = x^4 - 3 x^2 +1, \ldots$ In order to prove the result, it suffices to check the following relation:
\begin{eqnarray*}
(k+1) {d \over dt} \left( {J_{k+1} (2t) \over t} \right) 
- k {J_k (2t) \over t } = - (k+2) {J_{k+2} (2t) \over t }.
\end{eqnarray*}
The left hand side of this equation becomes
\begin{eqnarray*}
&&
(k+1) {2 J_{k+1} (2t) \over t} - (k+1) {J_{k+1} (2t) \over t^2 }- k {J_k (2t) \over t }
\\
&=&
{J_{k} (2t) \over t} - (k+1) {J_{k+1} (2t) \over t^2 }- (k+1) {J_k (2t) \over t }
\\
&=&
- (k+2) {J_{k+2} (2t) \over t },
\end{eqnarray*}
since the first and second equalities are obtained from recurrence formulas for the Bessel coefficients: $2 J_{k+1}'(2t)=J_k (2t) - J_{k+2}(2t)$ and $J_{k} (2t)+J_{k+2}(2t)=(k+1)J_{k+1}(2t)/t$ (see (1) in page 17 of \cite{wat}), respectively. This finishes the proof of the theorem.
\end{proof}

\section{A new type of limit theorems}

We can now state the main result of this paper. To do so, let define
\begin{eqnarray*}
| \tilde{\Psi}_{\infty} ^{(\infty)} (V_k ^{(\infty)},t) \rangle 
= \lim_{p \to \infty} \left\langle \Phi^{(p)} _k \left| \exp \left( i t {H^{(p)} _{\infty} \over \sqrt{p}}  \right) \right| \Phi^{(p)} _0 \right\rangle ,
\end{eqnarray*}
and
\begin{eqnarray*}
\tilde{P}_{\infty} ^{(\infty)} (V_k ^{(\infty)},t)
&=&
\langle \tilde{\Psi}_{\infty} ^{(\infty)} (V_k ^{(\infty)},t) | \tilde{\Psi}_{\infty} ^{(\infty)} (V_k ^{(\infty)},t) \rangle. 
\end{eqnarray*}
By Theorem 1 and the definition of $| \tilde{\Psi}_{\infty} ^{(\infty)} (V_k ^{(\infty)},t) \rangle$, we see that
\begin{eqnarray}
\sum_{k=0} ^{\infty} \tilde{P}_{\infty} ^{(\infty)} (V_k ^{(\infty)},t)
= \sum_{k=1} ^{\infty} k^2 \> {J_k ^2 (2t) \over t^2} = 1.
\label{eqn:arakawa}
\end{eqnarray}
The second equality comes from an expansion of $z^2$ as a series of squares of Bessel coefficients (see page 37 in \cite{wat}):
\begin{eqnarray*}
z^2 = 4 \sum_{k=1} ^{\infty} k^2 \> J_k ^2 (z). 
\end{eqnarray*}
Noting the result (\ref{eqn:arakawa}), here we define another continuous-time quantum walk $Y (t)$ starting from the root defined by 
\begin{eqnarray*}
P(Y (t) = k) =  
\tilde{P}_{\infty} ^{(\infty)} (V_k ^{(\infty)},t)  = (k+1)^2 \> {J_{k+1} ^2 (2t) \over t^2}.
\end{eqnarray*}
Therefore we obtain
\begin{theorem} 
\begin{eqnarray*}
 {Y(t) \over t} \>\> \Rightarrow \>\> Z,
\end{eqnarray*}
as $t \to \infty$, where $\Rightarrow$ means the weak convergence and $Z$ has the following density function:
\begin{eqnarray*}
I_{(0,2)}(x) \> {x^2 \over \pi \sqrt{4 - x^2}}.
\end{eqnarray*}
\end{theorem}
\begin{proof} From Theorem 1, we begin with computing 
\begin{eqnarray*}
E \left( \exp \left( i \xi {Y(t) \over t} \right) \right)
&=&
{\exp (-i \xi/t) \over t^2} \> \sum_{k=1} ^{\infty} \exp \left( i \xi {k \over t} \right) k^2 J_{k+1} ^2 (2t),
\end{eqnarray*}
for $\xi \in \RM$. By Neumann's addition theorem (see p.358 in \cite{wat}), we have
\begin{eqnarray*}
J_0 (\sqrt{a^2 + b^2 - 2 ab \cos (\xi)}) = \sum_{k= -\infty} ^{\infty} J_k(a)J_k(b) \exp (ik \xi).
\end{eqnarray*}
Taking $t=a=b$ in this equation gives
\begin{eqnarray*}
J_0 ( 4t \sqrt{\sin (\xi/2)}) = \sum_{k= -\infty} ^{\infty} J_k ^2 (t) \exp (ik \xi).
\end{eqnarray*}
By differentiating both sides of the equation with respect to $t$ twice, we see   
\begin{eqnarray*}
\sum_{k=1} ^{\infty} k^2 J_k ^2 (t) \exp (ik \xi) 
&=& 
{1 \over 2} \sum_{k= -\infty} ^{\infty} k^2 J_k ^2 (t) \exp (ik \xi) 
\\
&=& 
{t \over 4} \sin \left( {\xi \over 2} \right) J_0 ' \left( 2t \sin \left( {\xi \over 2} \right) \right) - {t^2 \over 2} \cos ^2 \left( {\xi \over 2} \right) J_0 '' \left( 2t \sin \left( {\xi \over 2} \right) \right). 
\end{eqnarray*}
Therefore we obtain
\begin{eqnarray*}
E \left( \exp \left( i \xi {Y(t) \over t} \right) \right)
&=&
\exp \left( -{ i \xi \over t} \right)  \> \left\{ 
{1 \over 2t} \sin \left( {\xi \over 2t} \right) J_0 ' \left( 4t \sin \left( {\xi \over 2t} \right) \right) \right.
\\
&&
\left.
\qquad \qquad \qquad \qquad \qquad 
- 2 \cos ^2 \left( {\xi \over 2t} \right) J_0 '' \left( 4t \sin \left( {\xi \over 2t} \right) \right) \right\}.
\end{eqnarray*}
Then a similar argument in \cite{kon05} yields
\begin{eqnarray*}
\lim_{t \to \infty} E \left( \exp \left( i \xi {Y(t) \over t} \right) \right)
= -2 J_0 '' ( 2 \xi ).
\end{eqnarray*}
On the other hand, (\ref{eqn:trino}) with $\nu =0$ gives
\begin{eqnarray*}
J_0 '' ( 2 \xi ) = - \int_{- 1} ^{1} \> \exp \left( 2 i \xi x \right) \> {x^2 \over \pi \sqrt{1 - x^2}} \> dx. 
\end{eqnarray*}
From the last two equations, we conclude that
\begin{eqnarray*}
\lim_{t \to \infty} E \left( \exp \left( i \xi {Y(t) \over t} \right) \right)
= \int_0 ^2 \> \exp \left( i \xi x \right) \> {x^2 \over \pi \sqrt{4 - x^2}} \> dx. 
\end{eqnarray*}
\end{proof}

It is interesting to remark that when $p=2$ case, i.e., $\ZM^1$, a similar type of density function was derived from a limit theorem for $X(t)$ (see \cite{kon05}):
\begin{eqnarray*}
\lim_{t \to \infty} E \left( \exp \left( i \xi {X(t) \over t} \right) \right)
= \int_0 ^1 \> \exp \left( i \xi x \right) \>  { 2 \over \pi \sqrt{1 - x^2}} \> dx. 
\end{eqnarray*}

\par
\
\par\noindent
{\bf Acknowledgments}
\par\noindent
The author is grateful to Kenichi Mitsuda and Etsuo Segawa for helpful discussions.


\begin{small}

\end{small}

\end{document}